\documentstyle[12pt]{article}

\begin{document}
\baselineskip=20.5pt
\def\beqra{\begin{eqnarray}} \def\eeqra{\end{eqnarray}}
\def\beqast{\begin{eqnarray*}} \def\eeqast{\end{eqnarray*}}
\def\beq{\begin{equation}}      \def\eeq{\end{equation}}
\def\be{\begin{enumerate}}   \def\ee{\end{enumerate}}

\def\fnote#1#2{\begingroup\def\thefootnote{#1}\footnote{#2}\addtocounter
{footnote}{-1}\endgroup}



\def\gam{\gamma}
\def\Gam{\Gamma}
\def\la{\lambda}
\def\eps{\epsilon}
\def\La{\Lambda}
\def\si{\sigma}
\def\Si{\Sigma}
\def\al{\alpha}
\def\Th{\Theta}
\def\th{\theta}
\def\tnu{\tilde\nu}
\def\vphi{\varphi}
\def\del{\delta}
\def\Del{\Delta}
\def\ab{\alpha\beta}
\def\om{\omega}
\def\Om{\Omega}
\def\mn{\mu\nu}
\def\mun{^{\mu}{}_{\nu}}
\def\kap{\kappa}
\def\rsi{\rho\sigma}
\def\beal{\beta\alpha}
\def\til{\tilde}
\def\rta{\rightarrow}
\def\eqv{\equiv}
\def\nab{\nabla}
\def\pa{\partial}
\def\sit{\tilde\sigma}
\def\ul{\underline}
\def\indt{\parindent2.5em}
\def\nd{\noindent}
\def\rsi{\rho\sigma}
\def\beal{\beta\alpha}
\def\caa{{\cal A}}
\def\cb{{\cal B}}
\def\cac{{\cal C}}
\def\cd{{\cal D}}
\def\ce{{\cal E}}
\def\cf{{\cal F}}
\def\cg{{\cal G}}
\def\cah{{\cal H}}
\def\ci{{\cal I}}
\def\cj{{\cal{J}}}
\def\ck{{\cal K}}
\def\cl{{\cal L}}
\def\cm{{\cal M}}
\def\cn{{\cal N}}
\def\cO{{\cal O}}
\def\cp{{\cal P}}
\def\car{{\cal R}}
\def\cs{{\cal S}}
\def\ct{{\cal{T}}}
\def\cu{{\cal{U}}}
\def\cv{{\cal{V}}}
\def\cw{{\cal{W}}}
\def\cx{{\cal{X}}}
\def\cy{{\cal{Y}}}
\def\cz{{\cal{Z}}}
\def\asymptotic{{_{\stackrel{\displaystyle\longrightarrow}
{x\rightarrow\infty}}\,\, }} 
\def\asymptext{\raisebox{.6ex}{${_{\stackrel{\displaystyle\longrightarrow}
{x\rightarrow\pm\infty}}\,\, }$}} 
\def\epsilim{{_{\textstyle{\rm lim}}\atop
_{~~~\epsilon\rightarrow 0+}\,\, }} 
\def\Llim{{_{\textstyle{\rm lim}}\atop
_{~~~L\rightarrow \infty}\,\, }} 
\def\omegalim{{_{\textstyle{\rm lim}}\atop
_{~~~\omega^2\rightarrow 0+}\,\, }} 
\def\xlimp{{_{\textstyle{\rm lim}}\atop
_{~~x\rightarrow \infty}\,\, }} 
\def\xlimm{{_{\textstyle{\rm lim}}\atop
_{~~~x\rightarrow -\infty}\,\, }} 
\def\asymptoticp{{_{\stackrel{\displaystyle\longrightarrow}
{x\rightarrow +\infty}}\,\, }} 
\def\asymptoticm{{_{\stackrel{\displaystyle\longrightarrow}
{x\rightarrow -\infty}}\,\, }} 

\def\raisenot{\raise .5mm\hbox{/}}
\def\nota{\ \hbox{{$a$}\kern-.49em\hbox{/}}}
\def\notA{\hbox{{$A$}\kern-.54em\hbox{\raisenot}}}
\def\notb{\ \hbox{{$b$}\kern-.47em\hbox{/}}}
\def\notB{\ \hbox{{$B$}\kern-.60em\hbox{\raisenot}}}
\def\notc{\ \hbox{{$c$}\kern-.45em\hbox{/}}}
\def\notd{\ \hbox{{$d$}\kern-.53em\hbox{/}}}
\def\notbd{\ \hbox{{$D$}\kern-.61em\hbox{\raisenot}}} 
\def\note{\ \hbox{{$e$}\kern-.47em\hbox{/}}}
\def\notk{\ \hbox{{$k$}\kern-.51em\hbox{/}}}
\def\notp{\ \hbox{{$p$}\kern-.43em\hbox{/}}}
\def\notq{\ \hbox{{$q$}\kern-.47em\hbox{/}}}
\def\notW{\ \hbox{{$W$}\kern-.75em\hbox{\raisenot}}}
\def\notz{\ \hbox{{$Z$}\kern-.61em\hbox{\raisenot}}}
\def\notpa{\hbox{{$\partial$}\kern-.54em\hbox{\raisenot}}}
\def\fo{\hbox{{1}\kern-.25em\hbox{l}}}  
\def\rf#1{$^{#1}$}
\def\bx{\Box}
\def\tr{{\rm Tr}}
\def\rmtr{{\rm tr}}
\def\dgg{\dagger}
\def\lag{\langle}
\def\rag{\rangle}
\def\bmid{\big|}
\def\vlap{\overrightarrow{\La p}} 
\def\lrta{\longrightarrow} \def\lrar{\raisebox{.8ex}{$\longrightarrow$}}
\def\ON{{\cal O}(N)}
\def\UN{{\cal U}(N)}
\def\bdPh{\mbox{\boldmath{$\dot{\!\Phi}$}}}
\def\bPh{\mbox{\boldmath{$\Phi$}}}
\def\bPhs{\bPh^2}
\def\sef{S_{eff}[\sigma,\pi]}
\def\sigx{\sigma(x)}
\def\pix{\pi(x)}
\def\bph{\mbox{\boldmath{$\phi$}}}
\def\bphs{\bph^2}
\def\ex{\BM{x}}
\def\exs{\ex^2}
\def\xdot{\dot{\!\ex}}
\def\y{\BM{y}}
\def\ys{\y^2}
\def\ydot{\dot{\!\y}}
\def\pat{\pa_t}
\def\pax{\pa_x}
\def\hp{{\pi\over 2}}

\renewcommand{\thesection}{\arabic{section}}
\renewcommand{\theequation}{\thesection.\arabic{equation}}

\begin{flushright}
\today\\
\end{flushright}

\vspace*{.1in}
\begin{center}
  \Large{\sc The Response to a Perturbation in the Reflection Amplitude}
\normalsize

\vspace{36pt}
{\large Joshua Feinberg\fnote{*}{{\it e-mail: joshua@physics.technion.ac.il}}}
\\
\vspace{12pt}
{\small \em Department of Physics,}\\ 
{\small \em Oranim-University of Haifa, Tivon 36006, Israel}\fnote{**}
{permanent address}\\
{\small and}\\
{\small \em Department of Physics,}\\
{\small \em Technion - Israel Institute of Technology, Haifa 32000 Israel}

\vspace{.6cm}

\end{center}

\begin{minipage}{5.8in}
{\abstract~~~~~
We apply inverse scattering theory to calculate the functional derivative of 
the potential $V(x)$ and wave function $\psi(x,k)$ of a one-dimensional 
Schr\"odinger operator with respect to the reflection amplitude $r(k)$.}

\end{minipage}

\vspace{58pt}
\noindent Key Words: Schr\"odinger operator, inverse scattering theory, 
Gelfand-Levitan-Marchenko equation, soliton, SUSY QM. \\
PACS numbers: 02.30.Zz, 03.65.Db, 03.65.Nk, 02.30.Rz        

\vfill
\pagebreak

\setcounter{page}{1}

\section{Introduction}
\setcounter{equation}{0}
The computation of the effect of a perturbation in the potential $V(x)$ 
of a one-dimensional Schr\"odinger operator 
\beq\label{schrodinger}
H=-\pax^2 + V(x)\,,
\eeq
defined on the whole real line on the {\em continuous} part of the spectrum of 
$H$ is a standard (albeit non trivial) procedure in perturbation theory. For
example, the response of the reflection amplitude $r(k)$ at momentum $k$ 
(in a setting where there is source at $x=+\infty$) to an infinitesimal 
change in the potential is 
\beq\label{drtodv}
{\delta r(k)\over\delta V(x)} = {\left(t(k)\phi(x,k)\right)^2\over 2ik} \,.
\eeq
The derivation of this result is given in the Appendix. Here $t(k)$ is the 
transmission amplitude and $t(k)\phi(x,k)$ is the solution of the 
Schr\"odinger equation 
\beq\label{schrodingereq}
\left(-\pax^2 + V(x) -k^2\right)\Psi(x,k) = 0
\eeq
which satisfies the scattering boundary conditions 
of this problem:
\beqra\label{phi}
t(k)\phi(x,k) &=& t(k)e^{-ikx} + o(1)\,\quad\quad\quad\quad\quad\quad ~~ 
x\rightarrow -\infty\nonumber\\
t(k)\phi(x,k) &=& ~~~~~e^{-ikx} + r(k)e^{ikx} + o(1)\,\quad\quad
x\rightarrow \infty\,.
\eeqra
Obtaining the kernel inverse to (\ref{drtodv}), i.e., the response of the 
Schr\"odinger potential $V(x)$ (as well as the response of the wave function) 
to a change 
in the reflection amplitude $r(k)$, is a much more difficult problem, which 
we solve in this paper. 

The explicit formulas we derive in this work express the {\em local} response 
of the potential and wavefunctions (see (\ref{dvdr}) and 
(\ref{dpsidrfinal})) to a change in the reflection amplitude. Thus, our 
results add to the information which can be gleaned from the well-known trace 
identities of \cite{trace}, the lowest of which reads 
\beq\label{traceid}
\int\limits_{-\infty}^\infty\,V(x)
\,dx = -{2\over \pi}\,\int\limits_0^\infty\,\log\,\left[1-|r(k)|^2
\right]\,dk-4\sum_{l=1}^N \kappa_l\,,
\eeq
where $E_l = -\kappa_l^2$ are the $N$ bound state energies 
(with $\kappa_l>0$), which tell us only about that response integrated over 
space.

A possibly interesting application of the results of this paper might be 
the investigation of perturbations around {\em reflectionless} potentials
with arbitrary numbers of bound states\cite{reflectionless}, which play an
important role in supersymmetric quantum mechanics\cite{cooper} and the theory
of solitons.

It would be useful at this point to introduce some additional notations and 
recall some basic facts, which will be used later on. For $k$ real, $H-k^2$ is 
real, and therefore 
$\Psi^*(x,k) = \Psi(x,-k)$ for any solution of (\ref{schrodingereq}). 
It follows that 
\beq\label{rconjugation}
r^*(k) = r(-k)\quad\quad{\rm and}\quad\quad t^*(k)=t(-k)
\eeq
in (\ref{phi}). 

$\Psi(x,k)$ and $\Psi^*(x,k)$ are linearly independent solutions of 
(\ref{schrodingereq}), and the continuous spectrum is doubly degenerate at
each $k^2>0$. In particular, $\phi(x,k)$ and $\phi^*(x,k)$ form a basis. 
Since they are degenerate in energy, their Wronskian is a non-zero constant.
Equating its values at $x\rightarrow\pm\infty$, we obtain the probability 
conservation relation
\beq\label{unitarity}
|r(k)|^2 + |t(k)|^2 =1\,.
\eeq
An equally suitable basis is the pair of solutions $\psi(x,k)$ and
$\psi^*(x,k)$ of (\ref{schrodingereq}), in which $\psi(x,k)$ obeys the boundary
condition 
\beq\label{psibc}
\psi(x,k) = e^{-ikx} + o(1)\,,\quad\quad x\rightarrow \infty\,.
\eeq
Thus, $\psi^*(x,k)$ corresponds to a setting in which there is a source
at $x=-\infty$.

We see from (\ref{phi}) and (\ref{psibc}) that 
$$\phi(x,k)\asymptotic \frac{1}{t(k)}\psi(x,k) + \frac{r(k)}{t(k)}\psi^*(x,k)
\,.$$
This relation must hold for all $x$, since $\psi(x,k)$ and $\psi^*(x,k)$ form 
a basis everywhere. Adding to it the linear combination for $\phi^*(x,k)$, 
we may write the relation between the two bases as
\beqra\label{psiphi}
\left(\begin{array}{c} \phi(x,k) \\{}\\ \phi^*(x,k) \end{array}\right) = 
\left(\begin{array}{cc} {1\over t(k)} & 
{r(k)\over t(k)} \\{}&{}\\ {r^*(k)\over t^*(k)} & 
{1\over t^*(k)}\end{array}\right)
\left(\begin{array}{c} \psi(x,k)\\{}\\ \psi^*(x,k)\end{array}\right)\,.
\eeqra
Note that the transformation matrix has a unit determinant. The inverse 
transformation is thus
\beqra\label{phipsi}
\left(\begin{array}{c} \psi(x,k) \\{}\\ \psi^*(x,k) \end{array}\right) = 
\left(\begin{array}{cc} {1\over t^*(k)} & 
-{r(k)\over t(k)} \\{}&{}\\ -{r^*(k)\over t^*(k)} & 
{1\over t(k)}\end{array}\right)
\left(\begin{array}{c} \phi(x,k)\\{}\\ \phi^*(x,k)\end{array}\right)\,.
\eeqra

This paper is organized as follows: In the next section we present a lightning 
review of inverse scattering theory. In particular, we discuss the 
Gelfand-Levitan-Marchenko equation and its properties. We show that its
solution is simply the boundary column of its resolvent kernel.

In section 3 we compute the variational derivative of the solution of the 
Gelfand-Levitan-Marchenko equation with respect to the 
reflection amplitude. Then we derive from it the corresponding derivatives 
of the potential and wavefunctions (Eqs. (\ref{dvdr}) and (\ref{dpsidrfinal}),
respectively) in closed form.

In section 4 we demonstrate the consistency of our results by comparing their
integrated form against known facts. 

Finally, in the Appendix we provide some useful technical details. 
In particular, we present the derivation of (\ref{drtodv}), and also discuss 
briefly the case of reflectionless potentials.

\section{The Gelfand-Levitan-Marchenko Equation and Its Solution} 
\setcounter{equation}{0}
According to inverse scattering theory (IST)
\cite{faddeev,faddeev1,faddeev2,novikov}, a Schr\"odinger operator 
(\ref{schrodinger}) whose potential $V(x)$ tends 
asymptotically to zero fast enough, such that $\int\limits_{-\infty}^\infty
\,|V(x)| (1+|x|)\, dx <\infty$, and thus supports only a finite number $N$
of bound states, is uniquely determined by the so-called scattering data. 
The scattering data are the reflection amplitude $r(k)$, and a finite set of 
$2N$ real numbers 
\beq\label{parameters}
\kappa_1 > \kappa_2 > \cdots \kappa_N > 0\quad\quad  {\rm and}\quad\quad
c_1, c_2, \cdots c_N\,,
\eeq 
where $E_l = -\kappa_l^2$ is the $l$th bound state energy, and where $c_l$
appears in the asymptotic behavior of the $l$th {\em normalized} bound state 
wave function as $c_l\exp -\kappa_l|x|$ (and thus determines its ``center of 
gravity''). 

The prescribed reflection amplitude $r(k)$ can be taken as any
complex-valued function which satisfies (for $k$ real): 
\beqra\label{rconditions}
r(-k)&=&r^*(k)\nonumber\\ 
|r(k)|&<&1\,,\quad\quad\quad\quad\quad k\neq 0\quad\quad 
{\rm and}\nonumber\\
r(k)&=&O\left(\frac{1}{k}\right)\,,\quad\quad |k|\rightarrow\infty\,.
\eeqra 
The first two conditions were already mentioned in (\ref{rconjugation}) and 
(\ref{unitarity}), and the third one reflects the fact that $V(x)$ is a small 
perturbation at high energy. 
In addition to these conditions, there is a less obvious technical condition 
that the Fourier transform $B(x) = \int\limits_{-\infty}^\infty 
\frac{r(k)}{t(k)} e^{ikx} dk$, should satisfy the 
bound  $\int\limits_{-\infty}^\infty (1+|x|) \large|\frac{dB(x)}{dx}\large| 
dx < \infty$.
Note further that $t(k)$ is completely determined by $r(k)$ as
\beq\label{transmission}
{t(k)\over\sqrt{1-|r(k)|^2}} = \left(\prod_{l=1}^N {k+i\kappa_l
\over k-i\kappa_l}\right)\,\exp\left({1\over 2\pi i}{\rm P.P.}
\int\limits_{-\infty}^\infty\,{\log\,\left[1-|r(q)|^2\right]\over q-k}dq
\right)\,.
\eeq

Given the scattering data, IST instructs us to determine a certain 
real transformation kernel $K(x,y)$, bounded on the domain $y\geq x$, 
which maps the wave functions of the free Schr\"odinger operator 
$H_0 = -\pax^2$ onto those of the operator $H$ in (\ref{schrodinger}). 
For example, the left moving wave $e^{-ikx}$ is mapped onto 
\beq\label{psi}
\psi(x,k) = e^{-ikx} + \int\limits_x^\infty K(x,y)\, e^{-iky}\,dy\,,
\eeq
which is evidently the solution of (\ref{schrodingereq}) satisfying the 
boundary condition (\ref{psibc}) mentioned above. 
Finally, the potential $V(x)$ in (\ref{schrodinger}) is determined by $K(x,y)$ 
according to 
\beq\label{potential}
V(x) = -2\frac{d}{dx} K(x,x)\,.
\eeq

The kernel $K(x,y)$ is determined as the solution of the 
Gelfand-Levitan-Marchenko equation (GLM)
\cite{faddeev,faddeev1,faddeev2,novikov}, 
\beq\label{GLM}
K(x,y) + F(x+y) + \int\limits_x^\infty K(x,z)F(z+y)\,dz = 0\,,
\eeq
where the {\em real} function $F(x)$ is 
\beq\label{F}
F(x) = \sum_{l=1}^N c_l^2 e^{-\kappa_l x} + {1\over 2\pi}
\int\limits_{-\infty}^\infty\, r(k) e^{ikx}\,dk\,.
\eeq

For fixed $x$, (\ref{GLM}) is a Fredholm integral equation of the second 
type in the unknown function $\Phi(y) = K(x,y)$, 
\beq\label{fredholm}
\Phi(y) = f(y) + \lambda\int\limits_x^\infty N(y,z)\Phi(z)\,,
\eeq
with symmetric real kernel and given function 
\beq\label{kernel}
N(y,z) = N(z,y) = -F(y+z)\,,\quad\quad f(y) = -F(x+y) = N(y,x)
\eeq
respectively, and spectral parameter $\lambda = 1$. 

It is known that (\ref{GLM}) has a unique solution, i.e., the Fredholm 
determinant of (\ref{fredholm}) is not null at $\lambda = 1$ 
\cite{faddeev1,faddeev2}. It is easy to demonstrate this property in the 
case of reflectionless potentials (for which $r(k)=0$ for all $k$), as we show
in the Appendix. 

The unique solution of (\ref{fredholm}) at $\lambda = 1$ is given by 
\beq\label{solution}
\Phi(y) = f(y) + \int\limits_x^\infty R(y,z;1)f(z)\,dz\,,
\eeq
where  $R(y,z;\lambda)$ is the resolvent kernel of (\ref{fredholm}).
Note that for real values of $\lambda$, $R(y,z;\lambda)$ is manifestly 
real, when it exists.

It is useful at this point to introduce the operator $\hat N$ and the 
vectors $ |\Phi\rangle$ and $ |f\rangle$, which correspond to 
the kernel $N(y,z)$ and functions $\Phi(y)$ and $f(y)$. Thus, in 
obvious notations,  $N(y,z) = \langle y | \hat N |z \rangle 
= -F(y+z),\, \langle y  |f\rangle = f(y) = \langle y | \hat N 
 |x \rangle$ and  $\Phi(y)=  \langle y |\Phi\rangle$. 
Then, it is easy to see from (\ref{fredholm}) that 
\beq\label{solution1}
 |\Phi\rangle = {1\over {\bf 1} -\hat N} |f\rangle  = 
\frac{\hat N}{1-\hat N}\ |x\rangle\,. 
\eeq
Similarly, from (\ref{solution}) we deduce that 
$R(y,z;1) = \langle y | \hat R  |z \rangle$, where  
\beq\label{resolvent}
\hat R = \frac{1}{{\bf 1}-\hat N} - {\bf 1} = 
\frac{\hat N}{1- \hat N}\,. 
\eeq
Thus, by comparing (\ref{solution1}) and (\ref{resolvent}), we conclude that 
$ |\Phi\rangle  = \hat R  |x\rangle\,,$ i.e., 
\beq\label{Ksolution}
\Phi(y) = K(x,y) =  \langle y | \hat R |x \rangle\,.
\eeq
The solution of the GLM equation (\ref{GLM}) coincides with the $x$th column
of its resolvent kernel. It is manifestly a real function of $x$ and $y$. 

\section{The Variational Derivatives}
\setcounter{equation}{0}
In view of (\ref{Ksolution}), it is straightforward to compute the 
variation of $K(x,y)$ under small perturbations in $\hat N$. Thus, consider 
a perturbation $\hat N\rightarrow \hat N + \delta\hat N$, which induces the
variation $\hat R \rightarrow \hat R +  \frac{1}{{\bf 1}-\hat N} 
\delta\hat N \frac{1}{{\bf 1}-\hat N}$. Consequently $\frac{\delta R(y,z;1)}
{\delta N(a,b)} = \langle y | ({\bf 1}+\hat R) |a \rangle \langle b | 
 ({\bf 1}+\hat R) |z \rangle$. Thus, from (\ref{Ksolution})
\beq\label{variation}
{\delta K(x,y)\over \delta N(a,b)} = \left(\delta(y-a) + R(y,a;1)\right)
\, \left(\delta(b-x) + K(x,b)\right)\,.
\eeq
In this work we are interested in variations $\delta\hat N$ which 
result from a change $\delta r(k)$ in the reflection amplitude. Due to the 
first condition in (\ref{rconditions}) we must impose 
$\delta r(-k) = \delta r^*(k)$. Thus, with no loss of generality, we take the 
positive components $r(k), k\geq 0$ as the independent functional variables. 
Keeping that in mind, we obtain from (\ref{F}) and (\ref{kernel}) 
that 
\beq\label{dndr}
\frac{\delta N(a,b)}{\delta r(k)} = -{1\over 2\pi} e^{ik(a+b)}\,. 
\eeq
Thus, from (\ref{variation}) and (\ref{dndr}) we obtain
\beqra\label{dkdr1}
&&\frac{\delta K(x,y)}{\delta r(k)} = \int\limits_x^\infty \frac{\delta K(x,y)}
{\delta N(a,b)}\, \frac{\delta N(a,b)}{\delta r(k)}\,da\,db =\nonumber\\ 
&&-\frac{1}{2\pi}\,\left(e^{iky} + \int\limits_x^\infty\,R(y,a;1)\,e^{ika}\,da
\right) \left(e^{ikx} + \int\limits_x^\infty\,K(x,b)\,e^{ikb}\,db\right) \,.
\eeqra
From (\ref{psi}) and from the reality of $K(x,y)$ and $k$, we recognize the 
last factor in (\ref{dkdr1}) simply as $\psi^*(x,k)$. Thus, 
\beq\label{dkdr2}
\frac{\delta K(x,y)}{\delta r(k)} =  
-\frac{1}{2\pi}\,\left(e^{iky} + \int\limits_x^\infty\,R(y,a;1)\,e^{ika}\,da
\right) \psi^*(x,k)
\eeq
In order to simplify (\ref{dkdr2}) further, we have to study the function 
\beq\label{omega}
\omega(y,x;k) = \int\limits_x^\infty\,R(y,a;1)\,e^{ika}\,da\,.
\eeq
Observe from (\ref{solution}) (considered with a generic given function 
$f(y)$), that $\Omega(y,x;k) = e^{iky} + \omega(y,x;k)$ is the unique 
solution of the Fredholm equation $\Omega(y,x;k) = e^{iky} + 
\int\limits_x^\infty N(y,z) \Omega(z,x;k)\,dz$, from which we infer that 
$\omega(y,x;k)$ is the unique solution of 
\beq\label{omegaeq}
\omega(y,x;k) = -G(y,x;k) + \int\limits_x^\infty N(y,z) \omega(z,x;k)\,dz
\eeq
where
\beq\label{G}
G(y,x;k) = \int_x^\infty F(y+z) e^{ikz}\,dz\,. 
\eeq
Thus, $\pax\omega(y,x;k)$ satisfies
\beq\label{omega1}
\frac{\partial\omega(y,x;k)}{\partial x} = F(x+y)\left(e^{ikx} + 
\omega(x,x;k)\right)
 + \int\limits_x^\infty N(y,z) \frac{\partial\omega(z,x;k)}{\partial x}\,dz\,.
\eeq
From (\ref{omega}), (\ref{Ksolution}) and the fact that 
$R(y,z;1) = R(z,y;1)$, we obtain that $\omega(x,x;k) = \int\limits_x^\infty 
R(a,x;1) e^{ika}\,da = \int\limits_x^\infty K(x,a) e^{ika}\,da\,.$
Thus, the inhomogeneous term in (\ref{omega1}) is simply 
$F(x+y)\left(e^{ikx} + \int\limits_x^\infty 
K(x,a) e^{ika}\,da\right) = F(x+y)\psi^*(x,k) = - f(y)\psi^*(x,k)$, where we 
used (\ref{psi}). It is the given function $f(y)$ in (\ref{fredholm}) 
multiplied by a $y$-independent factor $-\psi^*(x,k)$. Thus, from linearity, 
the unique solution of (\ref{omega1}) is simply the solution of 
(\ref{fredholm}), multiplied by the same factor, namely, 
\beq\label{omega2}
\frac{\partial\omega(y,x;k)}{\partial x} = - K(x,y)\psi^*(x,k)\,.
\eeq
The initial condition for this equation at $x=y$ is obviously $\omega(y,y;k) = 
\int\limits_y^\infty K(y,a) e^{ika}\,da = \psi^*(y,k) - e^{iky}$. Thus, 
\beq\label{omegafinal}
\omega(y,x;k) = \int\limits_x^y \psi^*(z,k) K(z,y)\,dz + \psi^*(y,k) 
- e^{iky}\,.
\eeq
Substituting this result into (\ref{dkdr2}) we obtain our first main result:
\beq\label{dkdrfinal}
\frac{\delta K(x,y)}{\delta r(k)} =  
-\frac{1}{2\pi}\,\left(\psi^*(y,k) +  \int\limits_x^y \psi^*(z,k) K(z,y)\,dz 
\right)\psi^*(x,k)\,.
\eeq
Since $K(x,y)$ is a real kernel, we can write (\ref{dkdrfinal}) alternatively 
as
\beq\label{dkdrfinalconjugate}
\frac{\delta K(x,y)}{\delta r^*(k)} =  
-\frac{1}{2\pi}\,\left(\psi(y,k) +  \int\limits_x^y \psi(z,k) K(z,y)\,dz 
\right)\psi(x,k)\,.
\eeq
The formula for $\frac{\delta K(x,y)}{\delta r^*(k)}$ is the key for obtaining
the functional derivatives of the wave function $\psi(x,k)$ and potential 
$V(x)$ with respect to the reflection amplitude $r(k)$, since the former are
linear in $K(x,y)$. Thus, from  (\ref{potential}) and 
(\ref{dkdrfinalconjugate}) we obtain our second main result: 
\beq\label{dvdr}
\frac{\delta V(x)}{\delta r^*(k)} = -2\frac{d}{dx}\,
\frac{\delta K(x,x)}{\delta r^*(k)} = \frac{1}{\pi}\frac{d}{dx}\,\psi^2(x,k)
\,.
\eeq
Similarly, from (\ref{psi}) and (\ref{dkdrfinalconjugate}) we obtain 
\beq\label{dpsidrconjugate}
\frac{\delta\psi(x,k)}{\delta r^*(q)} =  
-\frac{1}{2\pi}\,\left[\int\limits_x^\infty e^{-iky}\left( \psi(y,q) +  
\int\limits_x^y \psi(z,q) K(z,y)\,dz \right)\,dy \right]\psi(x,q)\,.
\eeq
Reversing the order of integrations in the second integral according to 
$$\int\limits_x^\infty e^{-iky}\left(\int\limits_x^y \psi(z,q) K(z,y)\,dz 
\right)\,dy = \int\limits_x^\infty\,\psi(z,q)\left(\int\limits_z^\infty\, 
e^{-iky} K(z,y)\,dy\right)\,dz$$
and recognizing the $y$-integral on the right hand side of the last equation
as\newline $\int\limits_z^\infty\, e^{-iky} K(z,y)\,dy = \psi(z,k) - e^{-ikz}
\,$, we obtain from (\ref{dpsidrconjugate}) our third main result:
\beq\label{dpsidrfinal}
\frac{\delta\psi(x,k)}{\delta r^*(q)} =  
-\frac{1}{2\pi}\,\left(\int\limits_x^\infty \,\psi(z,q)\psi(z,k) 
\,dz \right)\psi(x,q)\,.
\eeq
It is interesting to note that both (\ref{dvdr}) and (\ref{dpsidrfinal})
are expressed purely in terms of the wave function $\psi(x,k)$. Note, however,
that (\ref{dvdr}) is local in $\psi$, whereas (\ref{dpsidrfinal}) is highly 
nonlocal. It would be interesting to interpret these features from a physical
point of view.

\section{Consistency Checks of (\ref{dvdr}) and (\ref{dpsidrfinal})}
\setcounter{equation}{0}
The integrated form of (\ref{dvdr}) should agree with the derivative 
$\frac{\delta}{\delta r^*(k)}\int\limits_{-\infty}^\infty V(x)\,dx $
obtained from the trace identity (\ref{traceid}). Note that (\ref{traceid})
is expressed purely in terms of the positive Fourier modes of $r(k)$, the 
independent functional variables in our problem. Thus, taking the derivative
of (\ref{traceid}), we obtain
\beq\label{fromtraceid}
\frac{\delta}{\delta r^*(k)}\,
\int\limits_{-\infty}^\infty V(x)\,dx
= -{2\over \pi}\,\frac{\delta}{\delta r^*(k)}\,\int\limits_0^\infty\,\log\,
\left[1-|r(q)|^2\right]\,dq = \frac{2}{\pi}{r(k)\over |t(k)|^2}\,,
\eeq
where we used $|t(k)|^2 = 1 - |r(k)|^2.$ 
This result should be confronted with the integrated form of (\ref{dvdr}). 
We obtain from the latter
\beq\label{fromdvdr}
\frac{\delta}{\delta r^*(k)}\,
\int\limits_{-\infty}^\infty V(x)\,dx
= {1\over \pi}\,\Llim\left[\psi^2(x,k)\right]_{-L}^L\,.
\eeq
From (\ref{psibc}) we observe that $\psi(L,k)\simeq e^{-ikL}$.
Similarly, from (\ref{phipsi}) and (\ref{phi}) we deduce the asymptotic 
behavior 
\beq\label{psiminusinfinity}
\psi(-L,k)\simeq \frac{1}{t^*(k)}e^{ikL} - \frac{r(k)}{t(k)} e^{-ikL}\,.
\eeq
Substituting (\ref{psiminusinfinity}) and $\psi(L,k)\simeq e^{-ikL}$
in (\ref{fromdvdr}), we obtain 
\beq\label{fromdvdr1}
\frac{\delta}{\delta r^*(k)}\,
\int\limits_{-\infty}^\infty V(x)\,dx = {1\over \pi}\,\Llim\left[
e^{-2ikL}\left(1-{r^2\over t^2}\right) - {1\over (t^*)^2}e^{2ikL}
+ {2r\over |t|^2}\right]\,,
\eeq
which coincides in the limit with (\ref{fromtraceid}) in the sense of 
distributions, since the first two rapidly oscillating terms on the right 
hand side of (\ref{fromdvdr1}), when smeared against any continuous bounded 
test function $U(k)$ will integrate to zero in the limit $L\rightarrow\infty$,
due to the Riemann-Lebesgue lemma. Thus, (\ref{dvdr}) has passed its first 
test.

A less trivial test of (\ref{dvdr}) arises from comparing the complex 
conjugate equation of (\ref{dvdr}), 
$\frac{\delta V(x)}{\delta r(k)} = \frac{1}{\pi}\frac{d}{dx}\,
\psi^{*2}(x,k)$ and (\ref{drtodv}). We see that one kernel is the inverse of 
the other: $\int\limits_{-\infty}^\infty\,\frac{\delta r(k)}{\delta V(x)}\,
\frac{\delta V(x)}{\delta r(q)}\,dx = \delta(k-q) \,.$
Thus, we must verify that
\beq\label{Gamma}
\Gamma(k,q) = {t^2(k)\over 2\pi ik}\int\limits_{-\infty}^\infty 
\phi^2(x,k){d\over dx}\psi^{*2}(x,q)\,dx
\eeq
is equal to $\delta(k-q)$. 
Using the identity $$\int\limits_{-\infty}^\infty\,F^2\frac{d}{dx}\,G^2\,
dx = \frac{1}{2}\left[\left(FG\right)^2\right]_{-\infty}^\infty +
\int\limits_{-\infty}^\infty\,FG\left(F\pax G - G\pax F\right)\,dx $$
with $F=\phi(x,k)$ and $G=\psi^*(x,q)$, we write (\ref{Gamma}) as 
\beq\label{Gamma1}
\Gamma(k,q) = {t^2(k)\over 2\pi ik}\left\{\frac{1}{2}\left[\left(\phi(x,k)
\psi^*(x,q)\right)^2\right]_{-\infty}^\infty +\int\limits_{-\infty}^\infty 
\phi(x,k)\psi^*(x,k)W(x,k,q)\,dx\right\}
\eeq
where
\beq\label{wronskian}
W(x,k,q) = \phi(x,k)\pax\psi^*(x,q) - \psi^*(x,q)\pax\phi(x,k)
\eeq
is the Wronskian of $\phi(x,k)$ and $\psi^*(x,q)$. Next, from the 
Schr\"odinger equation for these two functions it is easy to obtain the 
relation
\beq\label{wronskianrelation}
\phi(x,k)\psi^*(x,q)  = {1\over k^2-q^2}\pax W(x,k,q)\,.
\eeq
Substituting the last equation in (\ref{Gamma1}) we see that the integral
is given entirely by boundary terms as
\beq\label{Gamma2}
\Gamma(k,q) = {t^2(k)\over 4\pi ik}
\left[\left(\phi(x,k)\psi^*(x,q)\right)^2 + {1\over k^2-q^2}\,W^2(x,k,q)
\right]_{-\infty}^\infty \,.
\eeq
In order to proceed in the clearest and simplest way, we shall make a few 
observations. First, observe from (\ref{phi}), (\ref{psibc}) and 
(\ref{psiminusinfinity}) that the {\em difference} 
$\left(\phi(L,k)\psi^*(L,q)\right)^2 - \left(\phi(-L,k)\psi^*(-L,q)\right)^2$
is rapidly oscillating in the limit $L\rightarrow\infty$ for all $k,q$. Thus, 
as a distribution acting on bounded continuous functions of $k$ and $q$ it 
tends to zero. Thus, we should focus on the second term. When $k^2-q^2\neq 0$,
the difference $W^2(L,k,q)-W^2(-L,k,q)$ oscillates rapidly, and tends to zero, 
in the sense of distributions, in the limit $L\rightarrow\infty$, similarly 
to the first term. Recall that for $k^2=q^2$, i.e., when $\phi(x,k)$ 
and $\psi^*(x,q)$ correspond to the same energy, $W(x,k,q)$ is a constant, 
and thus cannot oscillate. In that case, however, the denominator in front of 
this term vanishes. Thus we should study the limit $q^2\rightarrow k^2$ 
carefully. 

Let us concentrate then on the region $k^2\simeq q^2$. Since we 
have taken 
the positive Fourier modes of $r(k)$ as the independent functional variables, 
it is enough to study the case $q\rightarrow k$. Thus, assuming 
$k-q=\epsilon $ with $|\frac{\epsilon}{k}|<<1$ and setting $k+q\simeq 2k$, we
obtain from (\ref{phi}), (\ref{psibc}) and (\ref{psiminusinfinity}) that 
\beqra\label{asymptoticwronskian}
W(L,k,k-\epsilon)&\simeq & {i\over t(k)}\left(2k e^{-i\epsilon L} - \epsilon
r(k) e^{2ikL}\right)\nonumber\\
W(-L,k,k-\epsilon)&\simeq &{i\over t(k)}\left(2k e^{i\epsilon L} - \epsilon
t(k)\frac{r^*(k)}{t^*(k)} e^{2ikL}\right)\,.
\eeqra
Substituting (\ref{asymptoticwronskian}) and $k^2-q^2\simeq 2k\epsilon$ 
in (\ref{Gamma2}), and recalling the first observation made right below 
(\ref{Gamma2}), we finally obtain 
\beq\label{Gammafinal}
\Gamma(k,q) = {\sin (2\epsilon L)\over \pi\epsilon} + {\rm R.O.T.}\,,
\eeq
where the acronym R.O.T. stands for the rapidly oscillating terms that 
tend to zero in the sense of distributions. The first term, tends of course to 
$\delta(k-q)$ as $L\rightarrow\infty$, as desired. (\ref{dvdr}) has passed 
the second check!

Our last consistency check concerns (\ref{dpsidrfinal}). The point is 
that for $x\rightarrow -\infty$, the integral on the right hand side of 
(\ref{dpsidrfinal}) is the orthogonality relation
\beq\label{psiorthogonality}
\int\limits_{-\infty}^\infty\,\psi(z,q)\psi(z,k)\,dz = 
{2\pi\over |t(k)|^2}\left(\delta(k+q)-r(k)\delta(k-q)\right)\,.
\eeq
Let us sketch the proof of (\ref{psiorthogonality}): 
in analogy with (\ref{wronskianrelation}) we deduce that 
\beq\label{wronskianrelation1}
\psi(x,k)\psi(x,q)  = {1\over k^2-q^2}\pax \tilde W(x,k,q)
\eeq
where 
\beq\label{tildewronskian}
\tilde W(x,k,q) = \psi(x,k)\pax\psi(x,q) - \psi(x,q)\pax\psi(x,k)
\eeq
is the Wronskian of the functions involved. Then, after integration, we obtain
$$\int\limits_{-L}^L\,\psi(z,q)\psi(z,k)\,dz = {1\over k^2-q^2}\left[
\tilde W(L,k,q) - \tilde W(-L,k,q)\right]$$
and consider this result as $L\rightarrow\infty$, using the asymptotic 
behavior (\ref{psibc}) and (\ref{psiminusinfinity}). For generic values of 
$k,q$, the difference of Wronskians is a rapidly oscillating function, which
as a distribution, tends to zero as $L\rightarrow\infty$. As in the previous 
discussion, we observe that when $k^2=q^2$, $\tilde W(x,k,q)$ is independent 
of $x$, and thus does not oscillate. By studying the limit $q\rightarrow \pm k$
carefully, we deduce (\ref{psiorthogonality}). 

On the left hand side of (\ref{dpsidrfinal}) we have 
\beq\label{leftside}
\frac{\delta\psi(-L,k)}{\delta r(-q)} = \frac{\delta}{\delta r(-q)}\left(
\frac{1}{t(-k)}e^{ikL} - \frac{r(k)}{t(k)} e^{-ikL}\right)\,,
\eeq
where we used (\ref{psiminusinfinity}). Then, from (\ref{transmission}) it 
follows that 
\beqra\label{dadr}
\frac{\delta t^{-1}(k)}{\delta r(p)} &=& {1\over 2t(k)}{r^*(p)\over |t(p)|^2}
\left[\delta(p-k) + \delta(p+k) + {1\over i\pi}{\rm P.P.}
\left({1\over p-k}-{1\over p+k}\right)\right]\nonumber\\
&=&{1\over 2\pi i t(k)}{r^*(p)\over |t(p)|^2}
\left({1\over p-k -i\epsilon}-{1\over p+k+i\epsilon}\right)\,.
\eeqra
Applying (\ref{dadr}) to (\ref{leftside}), and studying the resulting 
expression around $k\simeq q>0$, we find (keeping only the singular terms)
${1\over 2\pi\epsilon}{r(k)\over|t(k)|^2}\psi(-L,k)$, in accordance
with the right hand side of (\ref{dpsidrfinal}) and (\ref{psiorthogonality}).


\setcounter{equation}{0}
\setcounter{section}{0}
\renewcommand{\theequation}{A.\arabic{equation}}
\renewcommand{\thesection}{Appendix:}
\section{Miscellaneous Technical Details}
\vskip 5mm
\setcounter{section}{0}
\renewcommand{\thesection}{A}
\subsection{Perturbation Theory in the Continuum: Proof of (\ref{drtodv})}
Consider a perturbation $\delta V(x)$ of the Schr\"odinger equation 
(\ref{schrodingereq}). To make the problem well-defined, we assume that 
$\delta V(x)$ is localized around some point $x_0$. 

Under this perturbation, the solution of (\ref{schrodingereq}) will be 
shifted $\Psi(x)\rightarrow\Psi(x)+\delta\Psi(x)$, and our task is to 
compute $\delta\Psi(x)$ to first order in $\delta V(x)$. To this order,
we have to solve the equation 
\beq\label{linearizedeq}
(H-k^2) \delta\Psi(x,k) = -\delta V(x) \Psi(x,k)\,.
\eeq
The general solution is given in terms of the Green's function $G(x,y;k)$ 
of the operator $H-k^2$ (defined with the appropriate boundary conditions)
as
\beq\label{linearizedsolution}
\delta\Psi(x,k) = -\int\limits_{-\infty}^\infty
G(x,y;k)\delta V(y) \Psi(y,k)\, dy\,.
\eeq
Since this expression vanishes for $\delta V(y) =0$, there is no term
which is a solution of the homogeneous equation. 

The proper Green's function is the one which decays exponentially 
whenever one of its coordinate arguments tends to $\pm\infty$, when 
$k$ is lifted to the upper  half complex plane. There is a unique such Green's
function:
\beq\label{greens}
G(x,y;k) = 
{1\over w(k)}\left[\theta\left(x-y\right)\psi^*(x,k)\phi(y,k)
+\theta\left(y-x\right)\psi^*(y,k)\phi(x,k)\right]\
\eeq
where 
\beq\label{wronskian3}
w(k)= -W(x,k,k) = {2k\over it(k)}
\eeq
from (\ref{wronskian}). Thus, in particular, for $\Psi(x,k)=\phi(x,k)$ we
obtain 
\beq\label{linearizedsolution1}
\delta\phi(x,k) = -{1\over w}\left[\psi^*(x,k)\int\limits_{-\infty}^x
\phi(y,k)\,\delta V(y) \phi(y,k)\, dy\, + 
\phi(x,k)\int\limits_x^\infty\psi^*(y,k)\,\delta V(y) \phi(y,k)\, dy\right]\,.
\eeq
In the limit $x\rightarrow -\infty$, we see that 
\beq\label{deltaphiminusinfty}
\delta\phi(x,k)\rightarrow -{\phi(x,k)\over w}\int\limits_{-\infty}^\infty
\psi^*(y,k)\,\delta V(y) \phi(y,k)\, dy\,.
\eeq
Thus, the function 
\beq\label{tildephi}
\tilde\phi(x,k) = {\phi(x,k) + \delta\phi(x,k)\over 1 
-{1\over w}\int\limits_{-\infty}^\infty
\psi^*(y,k)\,\delta V(y) \phi(y,k)\, dy}
\eeq
tends to $e^{-ikx}$ as $x\rightarrow -\infty$, and should be identified 
with the ``$\phi$-function'' (\ref{phi}) of the perturbed potential $V(x) + 
\delta V(x)$. In (\ref{tildephi}) we should keep, of course, only terms 
up to linear order in $\delta  V(x)$. As $x\rightarrow\infty$, we see from 
(\ref{linearizedsolution1}), (\ref{phi}) and (\ref{psibc}) that 
$\tilde\phi(x,k)\rightarrow {1\over \tilde t (k)} e^{-ikx}  + 
{\tilde r (k)\over \tilde t (k)} e^{ikx}$, where $\tilde t (k)$ and 
$\tilde r (k)$ are the new scattering amplitudes, given by 
\beqra\label{tildedamplitudes}
{1\over \tilde t (k)} &=&{1\over t(k)}\left[1+ 
{1\over w}\int\limits_{-\infty}^\infty\psi^*(y,k)\,\delta V(y) \phi(y,k)\, dy
\right]\nonumber\\
{\tilde r (k)\over \tilde t (k)} &=&{r(k)\over t(k)}
\left[1+ 
{1\over w}\int\limits_{-\infty}^\infty
\left(\psi^*(y,k) - {t(k)\over r(k)}\phi(y,k)\right)
\,\delta V(y) \phi(y,k)\, dy\right]
\eeqra
from which we infer, after some simple algebra, involving (\ref{psiphi})
that 
\beq\label{deltar}
\delta r(k) = \tilde r(k) - r(k) = {1\over 2ik}
\int\limits_{-\infty}^\infty
\left( t(k)\phi(y,k)\right)^2\,\delta V(y)\, dy\,,
\eeq
from which (\ref{drtodv}) follows.

\subsection{Reflectionless Potentials}
We demonstrate in the following the positivity of the Fredholm determinant
of the GLM equation (\ref{GLM}) in the case of reflectionless potentials.

For reflectionless potentials, where $r(k)\equiv 0$,  
$N(y,z) =  -\sum_{l=1}^N c_l^2 e^{-\kappa_l (y+z)}$ in (\ref{fredholm}) is a 
degenerate kernel of finite rank $N$. Thus, the integral equation degenerates 
into a system of linear equations, and the Fredholm determinant becomes an 
$N$ dimensional determinant. For this reason, reflectionless potentials are 
so easy to treat within the formalism of IST. 

The Fredholm determinant (at $\lambda=1$) is given by 
$(c_1\cdots c_N)^2\det A$, where 
\beq\label{Amatrix}
A_{mn} = \delta_{mn} + \frac{c_mc_n}{\kappa_n+\kappa_m} 
e^{-(\kappa_n+\kappa_m)x}\,.
\eeq
This matrix is a member of a family of matrices of the general form
\beq\label{matrixfamily}
A^{(\nu)}_{mn} = \delta_{mn} + \frac{v_m v_n}{(\kappa_n+\kappa_m)^\nu}\,,
\eeq
where $v_n$ are the components of a real vector and $\nu$ a real number. 
(Recall also that all $\kappa_n>0$.) The matrix (\ref{Amatrix}) corresponds, 
of course to $v_n = c_n e^{-\kappa_n x}$ and $\nu=1$. 
It is easy to demonstrate that  (\ref{matrixfamily}) is a positive definite 
matrix for any $\nu\geq 0$. To prove this, it is enough to verify that 
$\xi^T A\xi > 0$ for any real vector $\xi$. Thus, consider 
\beqra\label{positive}
\xi^T A\xi &=& \xi^T\xi + \sum_{n,m}{\xi_n v_n\xi_m v_m\over 
(\kappa_n+\kappa_m)^\nu}\nonumber\\
&=&\xi^T\xi + {1\over \Gamma(\nu)}\,\int_0^\infty s^{\nu-1}
\left(\sum_{n=1}^N \xi_nv_n e^{-\kappa_n s}\right)^2\,ds \geq 0
\eeqra
manifestly. Thus, in particular, (\ref{Amatrix}) is positive definite
and so is the corresponding Fredholm determinant.

\end{document}